\def\Title{Quantum subroutine problem and the robustness of 
quantum complexity classes}
\def\Author{Harumichi Nishimura and Masanao Ozawa}
\documentstyle[12pt]{article}
  \let\ps=\psi
  \newtheorem{Theorem}{Theorem}[section]
  
  \newtheorem{Lemma}[Theorem]{Lemma}
  \newtheorem{Corollary}[Theorem]{Corollary}
  \newenvironment{Proof}{\begin{trivlist}
    \item[\hskip \labelsep {\em \indent Proof.}]}{\qed\end{trivlist}}
  \newcommand{\qed}{{\em QED}}

  \newcommand{\C}{{\bf C}}
  \newcommand{\N}{{\bf N}}

  \newcommand{\Z}{{\bf Z}}
  \renewcommand{\l}{\langle}
  \newcommand{\r}{\rangle}

  \newcommand{\cC}{{\cal C}}
  \newcommand{\cD}{{\cal D}}

  \newcommand{\cH}{{\cal H}}

  \newcommand{\al}{\alpha}                                             %
  \newcommand{\be}{\beta}                                              %

  \newcommand{\de}{\delta}                                             %
  \newcommand{\ep}{\epsilon}
  
  \newcommand{\ga}{\gamma}                                             %

  \newcommand{\mb}{\mbox}

  \newcommand{\ph}{\phi}                                               %

  \newcommand{\si}{\sigma}                                             %
  \newcommand{\ta}{\tau}                                               %
  \newcommand{\Si}{\Sigma}                                          
 \renewcommand{\ep}{\varepsilon}

\newcommand{\ket}[1]{|#1\rangle}

  \title{\bf \Title}
  \author{\sc \Author\\
  \small\em CREST, Japan Science and Technology,
  \small\em  Graduate School of Human Informatics\\ 
  \small\em  Nagoya University, Chikusa-ku, Nagoya 464-8601, Japan}
  \date{}

\begin{document}

\maketitle

\begin{abstract} 
This paper positively solves the quantum subroutine problem 
for fully quantum oracles. The quantum subroutine problem 
asks whether a quantum computer with an efficiently computable oracle 
can be efficiently simulated by a non-oracle quantum computer. 
We extends the earlier results obtained by 
Bennett, Bernstein, Brassard, and Vazirani, 
and by Aharonov, Kitaev, and Nisan 
to the case where the oracle evaluates a unitary operator 
and the computer is allowed to be in the superposition 
of a query state and a non-query state during computation. 
We also prove the robustness of {\bf EQP}, {\bf BQP}, and {\bf ZQP} 
under the above general formulation, extending the earlier results 
on the robustness of {\bf BQP} shown by Bennett et al.\
\end{abstract}


\section{Introduction}

In computational complexity theory, an oracle is described 
as a device for computing some Boolean function $f$ 
at unit cost per evaluation. 
This allows us to formulate questions such as, 
``If we added the power of computing $f$ to a Turing machine, 
which functions could be efficiently computed by that Turing machine?'' 
Many researchers have investigated the computational power 
of a quantum Turing machine (QTM) with an oracle 
which computes a Boolean function. 
Berthiaume and Brassard \cite{BB92} constructed an oracle relative to 
which the QTM is exponentially more efficient than 
any deterministic Turing machine, recasting the promise problem of 
Deutsch and Jozsa \cite{DJ92}. 
Bernstein and Vazirani \cite{BV97} subsequently constructed an oracle 
which produces a superpolynomial gap between the quantum and 
probabilistic Turing machines. This result was improved 
by Simon \cite{Sim97}, who constructed an oracle 
which produces an exponential gap between the quantum and 
probabilistic Turing machines. 
Extending Simon's idea and using some new techniques, 
Shor \cite{Sho97} gave quantum polynomial time algorithms 
for factoring problems and discrete logarithms. On the other hand, 
Bennett, Bernstein, Brassard, and Vazirani (BBBV) \cite{BBBV97} 
showed that relative to an oracle chosen uniformly at random, 
with probability 1, {\bf NP}-complete problems cannot be solved 
by a QTM in polynomial time.  

The notion of oracles for quantum computers can be naturally 
extended to a device for carrying out a unitary operator $U$ 
at unit cost per evaluation. 
This allows us to formulate questions such as, 
``If we added the power of carrying out $U$ to a QTM, 
which functions could be efficiently computed by that QTM?''  
If $U$ is efficiently carried out, an oracle QTM with $U$ 
seems no more powerful than a non-oracle QTM. 
In fact, in the classical case, 
if a language $L$ is efficiently computable, 
a non-oracle Turing machine can efficiently simulate 
an oracle Turing machine with $L$  
by substituting a machine computing $L$ for a query to $L$. 
However, in the case of quantum computing, 
we need to consider a superposition of a query state and a non-query state. 
Moreover, quantum states with query strings of different lengths 
may superpose, even if each element of the superposition 
is a query state. 
In these cases, if we merely substitute a QTM computing $U$ 
for a query to $U$, quantum coherence will collapse. 
Thus, in this paper, we discuss the following problem. 
If a unitary transformation $U$ is efficiently computable by a QTM, 
is there a QTM efficiently simulating an oracle QTM with $U$? 
This problem is called the {\em quantum subroutine problem}. 
BBBV \cite{BBBV97} solved the quantum subroutine problem 
in the case where an oracle evaluates a deterministic function 
and the machine enters a query state deterministically. 
Aharonov, Kitaev, and Nisan \cite{AKN98} solved 
the problem for quantum circuits instead of QTMs 
in the case where an oracle evaluates a probabilistic function. 
We will positively solve the quantum subroutine problem 
for fully quantum oracles, extending their results to the case 
where the oracle evaluates a unitary operator 
and the computer is allowed to be in the superposition of a query state 
and a non-query state during computation. 
We can solve the quantum subroutine problem 
using the simulation of QTMs by quantum circuits \cite{Yao93} 
(See \cite{NO99} more formally), generalized quantum controls \cite{Kit97}, 
and the simulation of quantum circuits by QTMs. 
However, we use a quantum analog of a time constructible function. 
Because, this method is simple and 
it can reduce the polynomial slowdown caused by inserting subroutines 
as much as possible, comparing with the method of using quantum circuits.    

For a complexity class $\cC$, we say that $\cC$ is {\em robust} 
if it holds the relation $\cC^\cC=\cC$. 
In computational complexity theory, it is known 
that the complexity classes {\bf P}, {\bf BPP}, 
and {\bf ZPP} are robust, while it remains still open whether 
several classes such as {\bf NP} and {\bf RP} are robust or not. 
In this paper, we investigate the robustness of 
the quantum complexity classes {\bf EQP}, {\bf BQP}, and {\bf ZQP}, 
the quantum analogs of {\bf P}, {\bf BPP}, 
and {\bf ZPP}, extending the earlier result due to BBBV \cite{BBBV97}, 
who showed the robustness of {\bf BQP} in the case 
where the machine enters a query state deterministically. 
Using a solution for the subroutine problem and 
the method of the proof of BBBV, we can show that 
{\bf EQP} and {\bf BQP} are robust in the general case 
where a query state and a nonquery state may superpose. 
By the method of BBBV, a query step of an oracle QTM can be replaced 
by a Monte Carlo non-oracle QTM, but their method does not work 
for Las Vegas algorithms. 
In order to prove the robustness of {\bf ZQP}, 
we improve their method by keeping a witness 
to distinguish the case where a QTM queries an oracle correctly 
from other cases. 

This paper is organized as follows. In Section 2 
we give definitions and basic theorems on QTMs. 
In Section 3 we introduce a stationary time constructible function, 
and solve the quantum subroutine problem by using this function. 
This section also contains the rigorous formulation of oracle QTMs. 
In Section 4 we show that {\bf EQP}, {\bf BQP}, and {\bf ZQP} 
are robust in general form, improving the method of BBBV\ 
and using a solution of the quantum subroutine problem.

\section{Preliminaries}
A quantum Turing machine (QTM) is a quantum system 
consisting of a processor, a bilateral infinite tape, 
and a head to read and write a symbol on the tape. 
The formal definition of a QTM as a mathematical structure 
is given as follows. 
A {\em processor configuration set} is a finite set 
with two specific elements $q_0$ and $q_f$, 
where $q_0$ represents the {\em initial processor configuration} 
and $q_f$ represents the {\em final processor configuration}. 
A {\em symbol set} is a finite set of the cardinality at least 2 
with a specific element denoted by $B$ and called the {\em blank}. 
A {\em Turing frame} is a pair $(Q,\Si)$ of 
a processor configuration set $Q$ and a symbol set $\Si$. 
In what follows, let $(Q,\Si)$ be a Turing frame. 
A {\em tape configuration} from a symbol set $\Si$ is 
a function $T$ from the set $\Z$ of integers to $\Si$ 
such that $T(m)=B$ except for finitely many $m\in\Z$. 
The set of all the possible tape configurations 
is denoted by $\Si^{\#}$. 
The {\em configuration space} of $(Q,\Si)$ 
is the product set $\cC(Q,\Si)=Q\times\Si^{\#}\times\Z$. 
A {\em configuration} of $(Q,\Si)$ is an element $C=(q,T,\xi)$ 
of $\cC(Q,\Si)$. 
Specifically, if $q=q_0$ and $\xi=0$ then $C$ is called 
an {\em initial configuration} of $(Q,\Si)$, 
and if $q=q_f$ then $C$ is called a {\em final configuration} of $(Q,\Si)$. 
The {\em quantum state space} of $(Q,\Si)$ is the Hilbert space $\cH(Q,\Si)$ 
spanned by $\cC(Q,\Si)$ with the canonical basis $\{\ket{C}|C\in\cC(Q,\Si)\}$ 
called the {\em computational basis}. 
A {\em quantum transition function} for $(Q,\Si)$ 
is a function from $Q\times\Si\times Q\times\Si\times\{-1,0,1\}$ 
into the complex number field $\C$. 
A {\em (single tape) prequantum Turing machine} is 
defined to be a triple $M=(Q,\Si,\de)$ consisting of a Turing frame 
$(Q,\Si)$ and a quantum transition function $\de$ for $(Q,\Si)$.

Let $M=(Q,\Si,\de)$ be a prequantum Turing machine.   
An element of $Q$ is called a {\em processor configuration} of $M$, 
the set $\Si$ is called the {\em alphabet} of $M$, 
the function $\de$ is called the {\em quantum transition function} of $M$, 
and an (initial or final) configuration of $(Q,\Si)$ is called 
the {\em (initial or final) configuration} of $M$. 
A unit vector in $\cH(Q,\Si)$ is called a {\em state} of $M$. 
The {\em evolution operator} of $M$ is 
a linear operator $M_\de$ on $\cH(Q,\Si)$ such that 
\begin{equation}\label{eq:dom} 
M_\de\ket{q,T,\xi}=\sum_{p\in Q,\ta\in\Si,d\in\{-1,0,1\}}
\de(q,T(\xi),p,\ta,d)\ket{p,T_\xi^\ta,\xi+d}
\end{equation}
for all $(q,T,\xi)\in\cC(Q,\Si)$, 
where $T_\xi^\ta$ is a tape configuration defined by 
$$ 
T_\xi^\ta(m)=\left\{
\begin{array}{ll}
\ta &\mbox{if}\ m=\xi,\\
T(m)&\mbox{if}\ m\neq \xi.
\end{array}\right.
$$ 
Eq.\ (\ref{eq:dom}) uniquely defines the bounded operator $M_\delta$ 
on the space ${\cal H}(Q,\Sigma)$ \cite{ON98}. 
A (single tape) prequantum Turing machine 
is said to be a {\em (single tape) quantum Turing machine (QTM)} 
if the evolution operator is unitary. 

The following theorem proved in \cite{ON98} characterizes 
the quantum transition functions that give rise to QTMs. 
If it is assumed that the head must move either 
to the right or to the left at each step, 
condition (c) of Theorem \ref{th:21} is automatically satisfied. 
In this case, Theorem \ref{th:21} is reduced to the result 
due to Bernstein and Vazirani \cite{BV97}.

\begin{Theorem}\label{th:21} 
A prequantum Turing machine $M=(Q,\Si,\de)$ is a QTM 
if and only if $\de$ satisfies the following condition.

{\rm (a)} For any $(q,\si)\in Q\times\Si$,
$$ 
\sum_{p\in Q,\ta\in\Si,d\in\{-1,0,1\}}|\de(q,\si,p,\ta,d)|^2=1.
$$

{\rm (b)} For any $(q,\si),(q',\si')\in Q\times \Si$ 
with $(q,\si)\neq(q',\si')$,
$$ 
\sum_{p\in Q,\ta\in\Si,d\in\{-1,0,1\}}\de(q',\si',p,\ta,d)^{*}
\de(q,\si,p,\ta,d)=0. 
$$

{\rm (c)} For any $(q,\si,\ta),(q',\si',\ta')\in Q\times \Si^2$,
$$ 
\sum_{p\in Q,d=0,1}\de(q',\si',p,\ta',d-1)^{*}\de(q,\si,p,\ta,d)=0. 
$$

{\rm (d)} For any $(q,\si,\ta),(q',\si',\ta')\in Q\times \Si^2$,
$$ 
\sum_{p\in Q}\de(q',\si',p,\ta',-1)^*\de(q,\si,p,\ta,1)=0.
$$
\end{Theorem}

Let $S\subseteq Q\times\Sigma$. 
A complex-valued function on $S\times Q\times\Si\times\{-1,0,1\}$ 
is {\em unidirectional}, if we have $d=d'$ 
whenever $\delta(p,\sigma,\tau,q,d)$ and $\delta(p',\sigma',\tau',q,d')$ 
are both non-zero, where $q\in Q$, $(p,\sigma),(p',\sigma')\in S$, 
$\tau,\tau'\in\Sigma$, and $d,d'\in\{-1,0,1\}$.  
A prequantum Turing machine (or QTM) is said to be {\em unidirectional} 
if the quantum transition function is unidirectional. 
It is easy to see that a unidirectional prequantum Turing machine 
is a unidirectional QTM 
if the quantum transition function is unidirectional. 
It is easy to see that a unidirectional prequantum Turing machine 
is a unidirectional QTM if it satisfies 
conditions (a) and (b) of Theorem \ref{th:21}. 
We can show the following lemma for a unidirectional QTM 
by a way similar to \cite{BV97}. This lemma allows us 
to extend a partially defined unidirectional quantum transition function 
so that it can characterize a QTM.

\begin{Lemma}[completion lemma]\label{th:22} 
Let $\delta'$ be a unidirectional function on 
$S\times Q\times\Sigma\times\{-1,0,1\}$, 
where $S\subseteq Q\times\Sigma$. 
Assume that $\delta'$ satisfies 
the following conditions {\rm (a)} and {\rm (b)}.

{\rm (a)} For any $(q,\si)\in S$,
$$
\sum_{p\in Q,\tau\in\Si,d\in\{-1,0,1\}}|\delta'(q,\si,p,\ta,d)|^2=1.
$$

{\rm (b)} For any $(q,\si),(q',\si')\in S$ with $(q,\si)\neq(q',\si')$, 
$$
\sum_{p\in Q,\tau\in\Sigma,d\in\{-1,0,1\}}
\de'(q',\si',p,\ta,d)^*\de'(q,\si,p,\ta,d)=0.
$$ 
Then there is a unidirectional QTM $M=(Q,\Sigma,\delta)$ 
such that $\delta(p,\sigma,q,\tau,d)=\delta'(p,\sigma,q,\tau,d)$ 
whenever $\delta'(p,\sigma,q,\tau,d)$ is defined.
\end{Lemma}

We shall give a formal definition of simulation. 
Let $M=(Q,\Sigma,\delta)$ and $M'=(Q',\Sigma',\delta')$ be QTMs. 
Let $t$ be a positive integer and $\epsilon>0$. 
Let $e:{\cal C}(Q,\Sigma)\rightarrow {\cal C}(Q',\Sigma')$ 
be an injection computable in polynomial time, 
$d:{\cal C}(Q',\Sigma')\rightarrow {\cal C}(Q,\Sigma)$ 
a function computable in polynomial time satisfying $d\cdot e={\rm id}$, 
and $f$ a function from $\N^2$ to $\N$. 
We say that $M'$ {\em simulates} $M$ for $t$ steps 
with {\em accuracy} $\epsilon$ and {\em slowdown} $f$ 
(under the {\em encoding} $e$ and the {\em decoding} $d$), 
if for any $C_0\in{\cal C}(Q,\Sigma)$, we have
$$
 \sum_{C'\in{\cal C}(Q,\Sigma)}
\left| 
\left|\langle C'|M_\delta^t|C_0\rangle\right|^2-
 \sum_{C\in d^{-1}(C')}
\left|
\langle C|{M}^{f(t,\lceil\frac{1}{\epsilon}\rceil)}_{\delta'}| e(C_0)\rangle
\right|^2
\right|\le\epsilon. 
$$ 
If $f$ depends only on $t$ and 
the above relation is satisfied for $\epsilon=0$, 
we merely say that $M'$ simulates $M$ for $t$ steps with slowdown $f$.

Let $M=(Q,\Si,\de)$ be an $m$-track QTM. 
Then $\Si$ can be factorized as 
$\Si=\Si_1\times\Si_2\times\cdots\times\Si_m$ 
and $T\in\Si^\#$ can be written in the form $(T^1,T^2,\ldots,T^m)$, 
where $T^i\in\Si_i^\#$ for $i=1,\ldots,m$. 
The function $T^i$ is called an $i$-th track configuration. 
For a string $x=x_0x_1\cdots x_{k-1}$ of length $k$, 
we denote by ${\rm T}[x]$ a tape (or track) configuration such that 
${\rm T}[x](i)=x_i\ \ (0\le i\le k-1),\ B\ \ $(otherwise). 
For any tape configuration $T$, we will write $T=(T^1,\ldots,T^j)$ 
if $T=(T^1,\ldots,T^j,{\rm T}[\ep],\ldots,{\rm T}[\ep])$, 
where $\ep$ denotes the empty string. 
Henceforth, $(q,(T^1,\ldots,T^k),0)$ abbreviate 
a configuration $(q,(T^1,\ldots,T^k,{\rm T}[\ep],\ldots,{\rm T}[\ep]),
(0,\ldots,0))$. 
Let $E(\hat{\xi}=j)$, $E(\hat{q}=p)$, $E(\hat{T}=T_0)$ 
and $E(\hat{T^i}=T_0)$ be respectively projections on 
$\mb{span}\{\ket{q,T,j}|q\in Q,\ T\in\Si^\#\}$, 
$\mb{span}\{\ket{p,T,\xi}|$ $T\in\Si^\#,\ \xi\in\Z\}$, 
$\mb{span}\{\ket{q,T_0,\xi}|q\in Q,\ \xi\in\Z\}$ 
and $\mb{span}\{\ket{q,T,\xi}|q\in Q,\ T=
(T^1,\cdots,T_0,\cdots,T^m)\in\Si^\#,\ \xi\in\Z\}$. 
A QTM $M=(Q,\Si,\de)$ is said to be {\em stationary}, 
if given an initial configuration $C$, 
there exists some $t\in\N$ satisfying 
$||E(\hat{\xi}=0)E(\hat{q}=q_f)M_\de^{t}\ket{C}||^2=1$ and 
for all $s<t$ we have $||E(\hat{q}=q_f) M_\de^{s}\ket{C}||^2=0$. 
The positive integer $t$ is called the {\em computation time} of $M$ 
for input state $\ket{C}$, 
and $M_\de^t\ket{C}$ is called the {\em final state} of $M$ for $\ket{C}$. 
Specifically, if $\ket{C}=\ket{q_0,{\rm T}[x],0}$, 
the integer $t$ is called the computation time on input $x$. 
A {\em polynomial time} QTM is a stationary QTM such that 
the computation time on every input is a polynomial 
in the length of the input. 
It is easy to see that a polynomial time bounded QTM  
(i.e., a QTM whose computation time on every input 
is bounded by a polynomial in the length of the input) 
can be simulated by a polynomial time QTM with at most a polynomial slowdown. 
Moreover, we say that $M=(Q,\Si,\de)$ is in {\em normal form} 
if $\de(q_f,\si,q_0,\si,1)=1$ for any $\si\in\Si$. 
Henceforth, we shall consider only unidirectional stationary 
normal form QTMs, since such restricted QTMs are computationally 
equivalent to general QTMs independent of constraints 
on the error probability of algorithms \cite{NO99}. 

We have discussed solely single tape QTMs, but our arguments 
can be easily adapted to multi-tape QTMs. See \cite{ON98} 
for the formulation of the multi-tape QTMs.    

\section{Solution of the quantum subroutine problem}
A {\em stationary time constructible (ST-constructible)} QTM 
of a function $f:\N\rightarrow\N$ is defined to be a QTM 
such that if the initial state is $\ket{q_0,{\rm T}[x],0}$, 
then the final state is $\ket{q_f,{\rm T}[x],0}$ 
and that the computation time is $f(|x|)$, 
where $|x|$ denotes the length of $x$. 
A function $f:\N\rightarrow\N$ is said to be 
{\em stationary time constructible (ST-constructible)} 
if there exists a stationary time-constructible QTM of $f$.

\begin{Lemma}\label{th:31}
For any $k\ge 2$, there is an ST-constructible 
monic polynomial of degree $k$.
\end{Lemma} 

\begin{Proof} 
We show this theorem by induction on $k$. 
First, when $k=2$, we consider a two-track 
QTM $M_2=(Q,\Si\times\{B,1\},\de)$ satisfying 
the following transition rules, 
where $Q=\{q_0,q_1,\ldots,q_5,q_f\}$ and 
$\Sigma$ is an arbitrary symbol set. 
Henceforth, let $\si\in\Si\backslash\{B\}$ 
and let $s_i$ be an arbitrary symbol 
in the alphabet of the $i$-th track.

$$ 
\begin{array}{ll}
\delta(q_0,(\sigma,B),q_1,(\sigma,B),-1)=1,&
\delta(q_4,(\sigma,1),q_4,(\sigma,1),1)=1,\\
\delta(q_1,(B,B),q_2,(B,B),1)=1,& 
\delta(q_4,(s_1,B),q_5,(s_1,B),-1)=1,\\
\delta(q_2,(\sigma,B),q_2,(\sigma,1),1)=1,&
\delta(q_5,(\sigma,1),q_3,(\sigma,B),-1)=1,\\
\delta(q_2,(B,B),q_3,(B,B),-1)=1,& 
\delta(q_5,(B,B),q_f,(B,B),1)=1\\
\delta(q_3,(\sigma,1),q_3,(\sigma,1),-1)=1,& \\
\delta(q_3,(B,B),q_4,(B,B),1)=1&
\end{array}
$$

The above partially defined function $\de$ 
can be extended to be total by the completion lemma. 
Assuming that the input is written on the first track, 
$M_2$ implements the following steps.

Step 1. The head of $M_2$ changes each scanned symbol $B$ 
to the symbol $1$ on the second track 
with moving one cell to the right 
until it scans $B$ on the first track. 
If the head scans $B$ on the first track, 
it goes to the left until it scans $B$ on the first track again 
and then moves one cell to the right. 

Step 2. We iterate the following operation 
until the second track comes to be empty, 
where we say that the $i$-th track $T^i$ is {\em empty} 
if $T^i={\rm T}[\ep]$. The head goes to the right 
until it scans $B$ on the second track and 
then moves one cell to the left. 
Afterward, the head changes the scanned $1$ to $B$ 
on the second track and moves one cell to the left, 
goes to the left until it scans $B$ again on the second track, 
and moves one cell to the right if it scans $B$.

The computation time of $M_2$ is $(2n+4)+\sum_{i=0}^{n}(2i+2)=n^2+5n+6$.  
 
Next, we assume that there exists a QTM $M'$ such that 
the initial state and the final state are equal 
except for the processor configuration and that 
the computation time is a monic polynomial $p(n)$ of degree $k$. 
Then, we consider a QTM $M$ which implements the following steps.
     
Step 1. The head of $M$ changes each scanned symbol $B$ to the symbol $1$ 
on an auxiliary track with moving one cell to the right 
until it scans $B$ on the first track. 
If the head scans $B$ on the first track, 
it goes to the left until it scans $B$ on the first track again 
and then moves one cell to the right. 

Step 2. We iterate the following operation until 
the auxiliary track comes to be empty. 
Firstly, the head goes to the right until it scans $B$ 
on the auxiliary track and then moves one cell to the left. 
Secondly, the head changes the scanned $1$ to $B$ on the auxiliary track 
and moves one cell to the left, goes to the left 
until it scans $B$ again on the auxiliary track, 
and moves one cell to the right if it scans $B$. 
Thirdly, the machine runs $M'$.  
Lastly, the head goes to the right 
until it scans $B$ on the auxiliary track 
and afterward it goes to the left until it scans $B$ 
again on the auxiliary track.

We can construct a partially defined unidirectional 
quantum transition function implementing the above steps 
similar to the case $k=2$. Thus, we obtain 
the quantum transition function of $M$ by the completion lemma. 
The computation time of $M$ is 
$(2n+4)+\sum_{i=1}^{n}(c_1i+c_2+p(n))=np(n)+O(n^2)$, 
where $c_1$ and $c_2$ are constant positive integers. 
By induction hypothesis on $k$, the computation time of $M$ 
is a monic polynomial of degree $k+1$. 
Therefore, the proof is completed. 
\end{Proof}

It can be verified that the following lemma follows from Lemma \ref{th:31}.

\begin{Lemma}\label{th:32}
For any polynomial $p$ of degree $k$, there is 
an ST-constructible function $f$ 
such that $p+f$ is an ST-constructible (and monotone increasing) 
polynomial of degree $k$.
\end{Lemma}

\begin{Proof} 
For $k\ge 2$, there exists an ST-constructible 
monic polynomial of degree $k$ by Lemma \ref{th:31}. 
Moreover, it can be verified that $2n+4$, $3n+4$, 
and a constant function are ST-constructible. 
For example, we can provide an ST-constructible QTM of $3n+4$ 
whose quantum transition function $\de$ satisfies the following condition, 
where $\sigma$ is an arbitrary non-blank element 
in the alphabet of that QTM. 
$$ 
\begin{array}{ll}
\delta(q_0,\sigma,q_1,\sigma,-1)=1,& \delta(q_2,B,q_4,B,-1)=1,\\
\delta(q_1,B,q_2,B,1)=1,& \delta(q_4,\sigma,q_4,\sigma,-1)=1,\\
\delta(q_2,\sigma,q_3,\sigma,0)=1,&\delta(q_4,B,q_f,B,1)=1.\\
\delta(q_3,\sigma,q_2,\sigma,1)=1,& \\
\end{array} 
$$
Thus, any polynomial $p(n)=\sum_{j=0}^ka_jn^j$ of degree $k$ 
is written in the form
\begin{equation}\label{eq:1} 
p(n)=b_kf_k+b_{k-1}f_{k-1}+\ldots+b_2f_2+b_1(2n+4)+b_0(3n+4)+b_{-1},
\end{equation}
where $b_{k},\ldots,b_0,b_{-1}\in\Z$, 
and $f_k,\ldots,f_2$ are ST-constructible monic polynomials 
of degree $k,\ldots,2$, respectively. 
Now let $f_1=2n+4$, $f_0=3n+4$, and $f_{-1}=b_{-1}$. 
Let $\{g_1,\ldots,g_{l}\}=\{b_jf_j\ |\ b_j<0\}$ and 
$\{h_1,\ldots,h_{m}\}=\{b_jf_j\ |\ b_j\ge 0\}$. 
Then, from Eq.\ (\ref{eq:1}) we have 
\begin{equation}\label{eq:11}
p(n)-g_1-\ldots-g_l=d_1+\ldots+d_m.
\end{equation} 
We can see that the left hand side of Eq.\ (\ref{eq:11}) 
is an ST-constructible polynomial in the form $p+f$, 
where $f$ is ST-constructible. 
Moreover, it can be easily verified that Eq.\ (\ref{eq:11}) 
can be modified to an equation such that 
its left hand side is monotone increasing. 
\end{Proof}
 
BBBV \cite{BBBV97} defined an oracle quantum Turing machine 
as the following special QTM. An oracle quantum Turing machine 
has a special tape called an {\em oracle tape}. 
Its processor configuration set contains 
special elements $q_q$ and $q_a$, which are respectively called 
the {\em prequery} processor configuration and 
the {\em postquery} processor configuration. 
All cells of the oracle tape are blank except for 
a single block of non-blank cells. 
Given a language $L$ called an {\em oracle language}, 
this machine evolves as follows.

(1) If the processor configuration is $q_q$ and 
the string $(x,b)$ is written on the oracle tape, 
where $(x,b)\in\{0,1\}^*\times\{0,1\}$, 
the processor enters $q_a$ while the contents of the oracle tape 
change to $(x,b\oplus L(x))$ deterministically in a single step, 
where $\oplus$ denotes the exclusive-or.
   
(2) If the processor configuration is not $q_q$, 
then the machine evolves according to the quantum transition function.
  
Moreover, BBBV mentioned the notion of 
more general oracle quantum Turing machines, 
which has an {\em oracle unitary transformation} 
instead of an oracle language. 
Now we formulate a quantum Turing machine 
with an oracle unitary transformation, and give its elementary properties. 
We assume without loss of generality that the processor enters $q_q$ 
only when the head position of the oracle tape is zero 
and that an oracle unitary transformation are length-preserving, 
i.e., a state representing a string of length $n$ 
is transformed into a superposition of states representing strings 
of length $n$. 

Let $Q$ be a processor configuration set with $q_q$ and $q_a$, 
let $\Si$ be a symbol set, 
let $\de$ be a function from $(Q\backslash\{q_q\})\times\Si\times 
(Q\backslash\{q_a\})\times\Si\times\{-1,0,1\}$ to $\C$, 
and let $U$ be a unitary transformation such that 
$U\ket{x}\in\mb{span}\{\ket{z}|z\in\{0,1\}^n\}$ for any $x\in\{0,1\}^n$. 
Then $M=(Q,\Si,\de,U)$ is said to be 
an {\em oracle prequantum Turing machine (with $U$)}. 
The {\em evolution operator} of $M$ is defined 
to be a linear operator $U_M$ on $\cH(Q,\Si)$ such that
$$ 
U_M\ket{q,T,\xi}=
\left\{
\begin{array}{l}
\sum_{p\in Q\backslash\{q_a\},\ta\in\Si,d\in\{-1,0,1\}}
\de(q,T(\xi),p,\ta,d)\ket{p,T_\xi^\ta,\xi+d}\ \ \ (q\neq q_q)\\
\sum_{y\in\{0,1\}^{|x|}}\l y|U|x\r\ket{q_a,{\rm T}[y],0}
\ \ \ (q=q_q,T={\rm T}[x],\xi=0)\\
\ket{q_a,T,\xi}\ \ \ (\mb{otherwise}).
\end{array}\right.
$$
If $U_M$ is unitary, $M$ is said to be 
an {\em oracle quantum Turing machine (oracle QTM)}. 
Then we can obtain the following necessary and sufficient conditions 
by a way similar to the proof of Theorem \ref{th:21} \cite{ON98}.  

\begin{Theorem}\label{th:33}
An oracle prequantum Turing machine $M=(Q,\Si,\de,U)$ 
is an oracle QTM if and only if 
the following quantum transition function $\de'$ 
for $((Q\cup\{r\})\backslash\{q_q,q_a\},\Si)$ satisfies 
conditions {\rm (a)}--{\rm (d)} of Theorem \ref{th:21}. 
Here, $r$ is an element which is not in $Q$.
$$ 
\de'(q,\si,p,\ta,d)=
\left\{
\begin{array}{ll}
\delta(q_a,\sigma,q_q,\tau,d)&\ (q=p=r)\\ 
\delta(q_a,\sigma,q,\tau,d)&\ (q=r,\ p\neq r)\\ 
\delta(q,\sigma,q_q,\tau,d)&\ (p=r,\ q\neq r)\\
\delta(q,\sigma,p,\tau,d)&\ (q\neq r,\ p\neq r).
\end{array}\right.
$$
\end{Theorem}

Similarly we can define a multi-tape oracle QTM. 
For example, if $M$ is a $k$-tape oracle QTM 
and the state $\ket{\ps}$ of $M$ is 
$\ket{q_q,(T^1,\ldots,T^{k-1},{\rm T}[x]),
(d_1,\ldots,d_{k-1},0)}$, the state $U_M\ket{\ps}$ 
is defined to be 
$$
U_M\ket{\ps}=\sum_{y\in\{0,1\}^{|x|}}\l y|U|x\r
\ket{q_a,(T^1,\ldots,T^{k-1},{\rm T}[y]),
(d_1,\ldots,d_{k-1},0)}.
$$ 
Then, the $k$-th tape is called an oracle tape. 
We can consider an oracle QTM with a language $L$, 
defined by BBBV, to be a multi-tape oracle QTM 
with the unitary transformation $U_L$ 
such that $U_L\ket{x,b}=\ket{x,b\oplus L(x)}$ 
for all $(x,b)\in\{0,1\}^*\times\{0,1\}$. 
In what follows, we denote by $M^U$ (or $M^L$) 
an arbitrary oracle QTM with a unitary transformation $U$ 
(or a language $L$).
 
We introduce a notion necessary for a solution of 
the quantum subroutine problem. We denote by ${\cal D}(M,x)$ 
the set
$$
{\cal D}(M,x)=\{C\in{\cal C}(Q,\Sigma)|\exists
 s\le t\ [\langle C|M_\delta^s|q_0,{\rm T}[x],0\rangle\neq 0]\},
$$
where $t$ is the computation time of $M$ on input $x$. 
Let $M=(Q,\Si,\de)$ be a QTM and 
$M'=(Q',\Si_1\times\Si_2,\de')$ be a QTM 
such that $Q\times\Si\subseteq Q'\times\Si_1$. 
We say that $M'$ {\em carries out} $M$ {\em with slowdown $f$}, 
if there exists a function $f:\N\rightarrow\N$ 
such that for any input $x$ of $M$ and $C\in\cD(M,x)$ 
there exists some $T'\in\Si^\#_2$ (depending on $x$), 
and that 
$$
M_{\de'}^{f(|x|)}\ket{C}\ket{T'}
=\sum_{C'\in\cC(Q,\Si)}\l C'|M_\de|C\r \ket{C'}\ket{T'},
$$
where $\ket{C}\ket{T'}$ denotes $\ket{q,(T,T'),\xi}$ for $C=(q,T,\xi)$. 
It is easy to verify that if a QTM $M'$ can prepare 
a track configuration $T'$ satisfying the above condition 
in polynomial time and if $M'$ carries out a QTM $M$, 
then $M'$ simulates $M$ for any arbitrary steps. 
We can define analogous notions for multi-tape QTMs and oracle QTMs. 
For any QTM $M=(Q,\Sigma,\delta)$ and any $r\in Q$, 
we obtain the oracle QTM $M^U=(Q',\Sigma,\delta',U)$ 
with $Q'=((Q\backslash\{r\})\times\{0,1\})\cup\{q_q,q_a\}$ 
satisfying the following conditions.
$$ 
\begin{array}{l}
\delta'((q,0),\sigma,(p,1),\tau,d)=\delta(q,\sigma,p,\tau,d),\\
\delta'((q,1),\sigma,(q,0),\sigma,0)=1,\\ 
\delta'((q,0),\sigma,q_q,\tau,d)=\delta(q,\sigma,r,\tau,d),\\
\delta'(q_a,\sigma,(p,1),\tau,d)=\delta(r,\sigma,p,\tau,d),\\
\delta'(q_a,\sigma,q_q,\tau,d)=\delta(r,\sigma,r,\tau,d).
\end{array}
$$ 
In particular, if $U$ is the identity operator, 
then $M^U$ carries out $M$. 
Thus, we can consider a QTM to be a special case of an oracle QTM. 

We say that a unitary transformation $U$ 
is {\em polynomial time computable} by a QTM $M$, 
if the final state of $M$ for the initial state 
$\ket{q_0,{\rm T}[x],0}$ with $|x|=n$ is 
$$
\sum_{y\in\{0,1\}^n}\langle y|U|x\rangle\ket{q_f,{\rm T}[y],0}
$$ 
and the computation time of $M$ is a polynomial in $n$.

The following theorem gives us the positive answer 
for the quantum subroutine problem.   
\begin{Theorem}\label{th:34}
If a unitary transformation $U$ is polynomial time computable 
by a QTM $M$, there are a polynomial $p$ and 
a polynomial time QTM $M'$ such that $M'$ carries out 
a polynomial time oracle QTM $M^U$ with slowdown $p$.
\end{Theorem}

\begin{Proof} 
Let the quantum transition functions of $M$, $M'$, and $M^U$ 
be $\de$, $\de'$, and $\de^u$ respectively. 
Let the computation times of $M^U$ and $M$ 
be $f(n)$ and $g(n)$ respectively. 
By Lemma \ref{th:32} we can assume that 
$g$ is monotone increasing ST-constructible. 
Let $h=g\circ f+f$. Now we consider a QTM $M'$ 
which implements the following steps on input $x$, 
where $M'$ has three tapes and the third tape consists of two tracks.

Step 1. $M'$ writes $1^{h(|x|)}$ on the second tape.

Step 2. $M'$ carries out a single step of $M^U$ 
by the following steps 2.1--2.3.

Step 2.1. If the processor configuration is $q_q$ and 
$y$ is written on the third tape of $M'$, 
which corresponds to the oracle tape of $M^U$, 
then the head of the second tape goes to the right until it scans $B$. 
At the same time, $M'$ runs a QTM carrying out $M$ for $g(n)$ steps 
on the third tape. Afterward, the head of the third tape goes to the right 
while writing a special symbol $*$ on each cell of the second track. 
Here, let $q$ and $p$ be processor configurations of $M$, 
let $q_0$ and $q_f$ be respectively 
the initial and final processor configurations of $M$, 
and let $\si$ and $\ta$ be arbitrary elements in the alphabet of $M$. 
Moreover, throughout this proof, 
let $s_1$ be an arbitrary first tape symbol of $M'$, 
and let $s_3$ be an arbitrary first track symbol of the third tape of $M'$.
$$ 
\begin{array}{l}
\delta'(q_q,(s_1,1,\sigma),(q,1),(s_1,1,\tau),(0,1,d))
=\delta(q_0,\sigma,q,\tau,d),\\
\delta'((q,1),(s_1,1,\sigma),(p,1),(s_1,1,\tau),(0,1,d))
=\delta(q,\sigma,p,\tau,d),\ \ (q\neq q_0,q_f)\\  
\delta'((q_f,1),(s_1,1,\sigma),q_2,(s_1,1,\sigma),(0,1,-1))=1,\\
\delta'(q_2,(s_1,1,B),q_3,(s_1,1,B),(0,1,1))=1,\\         
\delta'(q_3,(s_1,1,s_3),q_3,(s_1,1,(s_3,*)),(0,1,1))=1.
\end{array}
$$
If the head of the second tape scans $B$, 
then the heads of the second and third tape move to the left. 
Specifically, the head of the third tape changes 
each scanned special symbol $*$ to $B$ while going to the left.
$$ 
\begin{array}{l} 
\de'(q_3,(s_1,B,s_3),q_4,(s_1,B,s_3),(0,-1,-1))=1,\\
\de'(q_4,(s_1,1,(s_3,*)),q_4,(s_1,1,s_3),(0,-1,-1))=1.
\end{array}
$$
If the head of the third tape scans $B$, 
then $M'$ carries out an ST-constructible QTM $M_{g+1}$ 
of the function $g+1$ on input $y$ after moving one cell to the right. 
Here, $q$ and $p$ are processor configurations of $M_{g+1}$, 
the symbols $\si$ and $\ta$ are arbitrary elements 
in the alphabet of $M_{g+1}$, and $\delta_{g+1}$ is 
the quantum transition function of $M_{g+1}$.
$$ 
\begin{array}{l} 
\delta'(q_4,(s_1,1,B),q_5,(s_1,1,B),(0,-1,1))=1,\\
\delta'(q_5,(s_1,1,\sigma),(q,6),(s_1,1,\tau),(0,-1,d))
=\delta_{g+1}(q_0,\sigma,q,\tau,d),\\
\delta'((q,6),(s_1,1,\sigma),(p,6),(s_1,1,\tau),(0,-1,d))
=\delta_{g+1}(q,\sigma,p,\tau,d)\ \ (q\neq q_0,q_f).
\end{array}
$$
If the head of the second tape scans $B$, 
then it moves one cell to the right 
and after three steps the processor enters $q_a$.
$$  
\begin{array}{l}  
\delta'((q_f,6),(s_1,B,s_3),q_7,(s_1,B,s_3),(0,1,0))=1,\\
\delta'(q_7,(s_1,1,s_3),q_8,(s_1,1,s_3),(0,0,0))=1,\\
\delta'(q_8,(s_1,1,s_3),q_9,(s_1,1,s_3),(0,0,0))=1,\\
\delta'(q_9,(s_1,1,s_3),q_a,(s_1,1,s_3),(0,0,0))=1.
\end{array}
$$

Step 2.2. If the processor configuration is not $q_q$, 
then the head of the second tape goes to the right. 
Here, $q$ and $p$ are processor configurations of $M^U$.
$$ 
\begin{array}{l}
\de'(q,(s_1,1,s_3),(q,2),(s_1,1,s_3),(0,-1,0))=1,\\
\de'((q,2),(s_1,B,s_3),(q,3),(s_1,B,s_3),(0,1,0))=1,\\
\de'((q,3),(s_1,1,s_3),(q,3),(s_1,1,s_3),(0,1,0))=1.
\end{array}
$$
If the head of the second tape scans $B$, 
then it goes to the left. 
$$ 
\begin{array}{l}  
\de'((q,3),(s_1,B,s_3),(q,4),(s_1,B,s_3),(0,-1,0))=1,\\
\de'((q,4),(s_1,1,s_3),(q,4),(s_1,1,s_3),(0,-1,0))=1.
\end{array}
$$
If the head of the second tape scans $B$ again, 
then it moves one cell to the right. 
Afterward, $M'$ carries out a single step of $M^U$ 
on the first and the third tapes 
while the head of the second tape stays during one step.
$$
\begin{array}{l}
\delta'((q,4),(s_1,B,s_3),(q,5),(s_1,B,s_3),(0,1,0))=1,\\
\delta'((q,5),(\sigma_1,1,\sigma_3),p,(\tau_1,1,\tau_3),(d_1,0,d_3))
=\delta^u(q,(\sigma_1,\sigma_3),p,(\tau_1,\tau_3),(d_1,d_3)).
\end{array}
$$

Step 2.3. If the processor configuration is $q_q$ and 
the query string is not correctly written on the third tape, 
then the head of the second tape goes to the right, 
goes to the left after it scans $B$, 
moves one cell to the right after it scans $B$ again, 
and $M'$ enters the postquery configuration 
without changing the tape configuration.

We can implement step 1 since the function $h$ 
is polynomial time computable by a QTM. 
In step 2.1, $M'$ must carry out $M$, since if $M'$ only simulates $M$, 
then $M'$ may leave extra information and computational paths 
with different extra information do not interfere. 
The partially defined function $\de'$ satisfies 
the unitary conditions of quantum transition functions 
of multi-tape QTMs \cite{ON98}, 
so that there exists a QTM implementing step 2 by the completion lemma. 
It is easy to see that the QTM implementing step 2 
carries out $M^U$ with slowdown $2h(|x|)+5$ and 
that the computation time of $M'$ is a polynomial in $n$. 
\end{Proof}

Using ST-constructible functions, we have reduced 
polynomial slowdown caused by the insertion of subroutines 
as much as possible. 
The degree of this polynomial slowdown is same as 
the case of deterministic or probablistic Turing machines.

\begin{Corollary}
If a unitary transformation $U$ is computable 
by a QTM $M$ in linear time, there is a QTM $M'$ 
which carries out a linear time oracle QTM $M^U$ in quadratic time.
\end{Corollary}
 
Next, we consider the bounded error version of 
the quantum subroutine problem (If a unitary transformation $U$ 
is efficiently computable by a QTM with any accuracy, 
is there an efficient QTM simulating an oracle QTM $M^U$ with any accuracy?). 

A function $f:\N^2\rightarrow\N$ is said to be 
{\em stationary time constructible (ST-constructible)} 
if there is a QTM satisfying the following condition: 
the final state of the QTM for the initial state 
$\ket{q_0,({\rm T}[x],{\rm T}[y]),0}$ is 
$\ket{q_f,({\rm T}[x],{\rm T}[y]),0}$ and 
the computation time is $f(|x|,|y|)$. 
Then the following lemma holds by a proof 
similar to Lemmas \ref{th:31} and \ref{th:32}. 

\begin{Lemma}\label{th:36}
For any polynomial $p(n,m)=\sum_{c=1}^k\sum_{d=1}^l a_{c,d}n^cm^d$ 
over $\Z$, where $a_{k,l}\neq 0$, there is an ST-constructible function 
$f(n,m)=\sum_{c=1}^k\sum_{d=1}^l b_{c,d}n^c m^d$ with $b_{k,l}\neq 0$ 
such that $p+f$ is ST-constructible.
\end{Lemma}

We say that a unitary transformation $U$ is 
{\em approximately polynomial time computable} by a QTM $M$ 
if the following conditions hold. 

(1) $U\ket{x}\in\mb{span}\{\ket{z}|z\in\{0,1\}^n\}$ 
for any $x\in\{0,1\}^n$. 

(2) There is a family of unitary transformations $\{U'_l\}$ 
such that the final state of $M$ for the initial state 
$\ket{q_0,{\rm T}[x,1^l],0}$ with $|x|=n$ is 
$$ 
\sum_{y\in\{0,1\}^*}\l y|U'_l|x\r\ket{q_f,{\rm T}[y,1^l],0}
$$ 
and $||U\ket{x}-U'_l\ket{x}||\le 1/2^l$ for any $x\in\{0,1\}^*$. 

(3) The computation time of $M$ is a polynomial in $n$ and $l$.\\ 
Let $M=(Q,\Si,\de)$ be a QTM and 
$M'=(Q',\Si_1\times\Si_2,\de')$ be a QTM 
such that $Q\times\Si\subseteq Q'\times\Si_1$. 
We say that $M'$ {\em carries out} $M$ 
{\em with bounded error and slowdown $f$}, 
if there exists a function $f:\N^2\rightarrow\N$ 
such that for any input $y=(x,1^l)$ of $M$, 
where $l\in\N$, and any $C\in\cD(M,y)$ 
there exists some $T'\in\Si^\#_2$ (depending on $y$), 
and that $||M_{\de'}^{f(|x|,l)}\ket{C}\ket{T'}-
(M_\de\ket{C})\otimes\ket{T'}||\le 1/2^l$. 
Now using Lemma \ref{th:36} the following theorem holds 
by a way similar to the proof of Theorem \ref{th:34}, 
and gives the positive answer for the bounded error version 
of the quantum subroutine problem of QTMs. 

\begin{Theorem}\label{th:37} 
If a unitary transformation $U$ is 
approximately polynomial time computable by a QTM $M$, 
for any $l\in\N$ there are a polynomial $p(n,l)$ and 
a polynomial time QTM $M'$ such that $M'$ carries out 
a polynomial time oracle QTM $M^U$ with bounded error and slowdown $p$.
\end{Theorem}

\section{Robustness of quantum complexity classes}

In this section, we identify a language $L$ with 
its characteristic function $c_L$, and we denote $c_L(x)$ by $L(x)$. 
We shall now define complexity classes for oracle QTMs. 
These definitions naturally extend 
the notion of complexity classes for QTMs \cite{BV97,NO99}. 
In what follows, we assume that the ranges of quantum transition functions 
are the polynomial time computable numbers.

We say that an oracle QTM $M$ {\em accepts} (or {\em rejects}) 
$x\in\{0,1\}^*$ {\em with probability} $p$ 
if the final state $\ket{\psi}$ of $M$ for the initial state 
$\ket{q_0,{\rm T}[x],0}$ satisfies
$$ 
||E(\hat{T^1}={\rm T}[x])E(\hat{T^2}={\rm T}[1])\ket{\psi}||^2=p
\ \ ({\rm or}\ ||E(\hat{T^1}={\rm T}[x])E(\hat{T^2}={\rm T}[0])
\ket{\psi}||^2=p).
$$
We say that $M$ {\em recognizes} a language $L$ with probability $p$ 
if $M$ accepts $x$ with probability at least $p$ for any $x\in L$ 
and rejects $x$ with probability at least $p$ for any $x\not\in L$. 
Moreover, we say that $M$ recognizes $L$ 
{\em with probability uniformly larger than} $p$, 
if there is a constant $0<\eta\le 1-p$ such that 
$M$ recognizes $L$ with probability $p+\eta$. 
A language $L'$ is in ${\bf BQP}^L$ (or ${\bf EQP}^L$) 
if there is a polynomial time oracle QTM $M^L=(Q,\Si,\de,U_L)$ 
that recognizes $L'$ with probability uniformly larger than 
$\frac{1}{2}$ (with probability 1). 
Then, $M^L$ is called a {\bf BQP}-machine (or {\bf EQP}-machine). 
A language $L'$ is in ${\bf ZQP}^L$ 
if there is a polynomial time QTM $M^L=(Q,\Si,\de,U_L)$ 
satisfying the following conditions: (1) $M^L$ recognizes $L$ 
with probability uniformly larger than $\frac{1}{2}$; 
(2) If $M$ accepts (rejects) $x$ with a positive probability, 
$M$ rejects (accepts) $x$ with probability $0$. 
Such a QTM $M^L$ is called a {\bf ZQP}-machine. 
For classes $\cC$ and $\cD$ of languages, 
let $\cC^\cD=\bigcup_{L\in\cD}\cC^L$. 
If $\cC^\cC=\cC$, the class $\cC$ is said to be {\em robust}.

We can apply Theorems \ref{th:34} and \ref{th:37} 
to the robustness of the quantum complexity classes 
{\bf EQP} and {\bf BQP}. If $L$ is in ${\bf EQP}$, 
then we can construct a polynomial time oracle QTM 
such that only the input $x$ and the answer $L(x)$ 
are written on the tape of the final state with probability 1 
by the method of Bennett in reversible computation \cite{Ben73}. 
In other words we can assume that an {\bf EQP}-machine 
has only one accepting configuration. 
His method is implemented in the following steps. 
We compute $L(x)$, copy $L(x)$ into an extra track, 
and carry out the reverse of the process of computing $L(x)$ 
in order to get rid of the scratch work. In the case of QTMs, 
reverse computation can be implemented by using 
the reversal lemma due to Bernstein and Vazirani \cite{BV97}. 
By the method of Bennett, we can see that for any $L\in{\bf EQP}$, 
a unitary transformation $U_L$ such that 
$U_L\ket{x,b}=\ket{x,b\oplus L(x)}$ is polynomial time computable. 
Thus, ${\bf EQP}$ is robust by Theorem \ref{th:34}.  

\begin{Theorem}
${\bf EQP}^{\bf EQP}={\bf EQP}$. 
\end{Theorem}

BBBV \cite{BBBV97} showed the following theorem, 
which ensures the use of a Monte Carlo quantum algorithm 
as a subroutine of another quantum algorithm. 

\begin{Theorem}\label{th:42} 
If a language $L$ is in ${\bf BQP}$, 
for any $l\in\N$ there is a QTM $M$ 
which recognizes $L$ with probability $1-1/2^l$ and 
has the following property {\rm (A):} The computation time 
of $M$ for $\ket{q_0,{\rm T}[x],0}$ is a polynomial in $|x|$ and $l$, 
and the final state is $\al\ket{q_f,T,0}+\ket{\psi}$, 
where $|\al|^2\ge 1-1/2^l$ and $T=({\rm T}[x],{\rm T}[L(x)])$.
\end{Theorem} 

{\em Remark.} The QTM $M$ obtained in the proof in \cite{BBBV97} 
is not always stationary. However, we can construct 
a stationary QTM with property (A) by using the construction 
of a universal QTM \cite{BV97,Yao93,NO99}.

Theorem \ref{th:42} guarantees that without loss of generality 
a {\bf BQP}-machine recognizing $L$ has a clean tape 
with only the input $x$ and the answer $L(x)$ 
with arbitrary large probability after computation. 
In other words we can assume that a {\bf BQP}-machine 
has only one accepting configuration. 
BBBV \cite{BBBV97} claimed that ${\bf BQP}$ is robust 
as the corollary of Theorem \ref{th:42}, 
since this theorem allows us to use a QTM recognizing 
an oracle language instead of the oracle itself. 
However, they considered the case where the machine enters 
a query state deterministically, i.e., 
they did not discuss the possibility that 
the coherence of different computation paths collapses 
by the insertion of a QTM recognizing an oracle language. 
We have already solved this problem by Theorem \ref{th:37}, 
so that we can show that ${\bf BQP}$ is robust 
in the general setting where a query state and 
a nonquery state may superpose.  

\begin{Theorem}
${\bf BQP}^{\bf BQP}={\bf BQP}$. 
\end{Theorem}
        
Now we consider the robustness of ${\bf ZQP}$. 
If we apply Theorem \ref{th:42} to a language in {\bf ZQP}, 
the obtained algorithm will not be Las Vegas. 
Thus, we need the following theorem, 
which means that we can also assume that 
a {\bf ZQP}-machine has only one accepting configuration. 

\begin{Theorem}\label{th:44} 
If a language $L$ is in ${\bf ZQP}$, 
for any $l\in\N$ there is a QTM $M$ 
which recognizes $L$ with probability $1-1/2^l$ and 
has the following property {\rm (B):} 
The computation time of $M$ for $\ket{q_0,{\rm T}[x],0}$ 
is a polynomial in $|x|$ and $l$, and 
the final state is $\al\ket{q_f,T,0}+\ket{\psi}$, 
where $|\al|^2\ge 1-1/2^l$, $T=({\rm T}[x],{\rm T}[L(x)])$, 
and $E(\hat{q}=q_f)E(\hat{T}^2={\rm T}[\star])\ket{\psi}=\ket{\psi}$. 
Here, we denote by $\star$ a special symbol of the second track of $M$. 
\end{Theorem}

\begin{Proof} 
For simplicity, we represent a state of a QTM by its track configurations. 
We denote by $\ket{T^1}\ket{T^2}\cdots\ket{T^n}$ 
a computational basis vector of a QTM 
such that for each $i=1,\ldots,n$, 
the $i$-th track configuration is $T^i$. 
Let $L\in {\bf ZQP}$. Then, we can assume that 
there is a {\bf ZQP}-machine $M'$ 
which recognizes $L$ with probability $1-1/2^{l+1}$ 
in time polynomial in the length of input and $l$. 
Let the final state of $M'$ for the initial state $\ket{x}$ be 
\begin{equation}\label{eq:2}
\sum_{w}\al(w)\ket{x}\ket{L(x)}\ket{w}
+\sum_{y,z}\sum_{v}\be(y,z,v)\ket{y}\ket{z}\ket{v},
\end{equation}
where $x$ and $L(x)$ denote ${\rm T}[x]$ and ${\rm T}[L(x)]$ respectively, 
the summations $\sum_{v}$ and $\sum_{w}$ are respectively taken 
over all the third track strings, and $\sum_{y,z}$ is taken over 
all the pairs $(y,z)$ of the first and second track strings 
such that $(y,z)\neq (({\rm T}[x],{\rm T}[0]),({\rm T}[x],{\rm T}[1]))$. 
Then, we have $\sum_{w}|\al(w)|^2\ge 1-1/2^{l+1}$. 
Now we consider a {\bf ZQP}-machine $M$ with seven tracks 
which implements the following steps.

Step 1. $M$ on input $x$ writes $1^{p_l(|x|)}$ 
between cell $1$ and cell $p_l(|x|)$, and 
$0^{p_l(|x|)}$ between cell $-1$ and cell $-p_l(|x|)$ 
of the seventh track. Here, $p_l(|x|)$ is the computation time 
of $M'$ on input $x$. 

Step 2. $M$ runs $M'$.

Step 3. $M$ respectively copies the first and second track strings 
to the fourth and fifth tracks.

Step 4. $M$ runs the reverse of $M'$. 
This step is implementable by the reversal lemma.

Step 5. If $x$ is respectively written on the first and fourth tracks, 
the symbol $0$ or $1$ is written on the fifth track, 
and other tracks are empty, then $M$ writes no symbol. 
Otherwise, $M$ writes a special symbol $\star$ 
in the cell 0 of the sixth track.

Step 6. If the first track string and the fourth track string are equal,  
then $M$ erases the fourth track string.
 
Step 7. If $\star$ is written in the cell 0 of the sixth track, 
then $M$ exchanges the contents of the second track 
for those of the sixth track. 
Otherwise, $M$ exchanges the contents of the second track 
for those of the fifth track.

Now we shall verify that the desired state is obtained 
after steps 1--7. The state of the system after step 2 
is represented by Eq.\ (\ref{eq:2}). 
After step 3 the system will evolve into the state
\begin{eqnarray*}
\lefteqn{\sum_{w}\alpha(w)\ket{x}\ket{L(x)}\ket{w}\ket{x}\ket{L(x)}
+\sum_{y,z,v}\beta(y,z,v)\ket{y}\ket{z}\ket{v}\ket{y}\ket{z}}
\quad \\
&=&\left(\sum_{w}\alpha(w)\ket{x}\ket{L(x)}\ket{w}
+\sum_{y,z,v}\beta(y,z,v)\ket{y}\ket{z}\ket{v}\right)\ket{x}
\ket{L(x)}\\
& & +\sum_{y,z,v}\beta(y,z,v)\ket{y}\ket{z}\ket{v}
(\ket{y}\ket{z}-\ket{x}\ket{L(x)}).
\end{eqnarray*} 
Since the unitary transformation $U$ implementing step 4 
is identical on the fourth and fifth tracks, 
there is a unitary transformation $U'$ such that
$$
U(\ket{T^1}\ket{T^2}\ket{T^3}\ket{T^4}\ket{T^5})
=(U'\ket{T^1}\ket{T^2}\ket{T^3})\otimes(\ket{T^4}\ket{T^5}),
$$
where $T^i$ is an arbitrary $i$-th track configuration for $i=1,\ldots,5$. 
Then, the state of the system is
\begin{eqnarray*}
 \lefteqn{\ket{x}\ket{B}\ket{B}\ket{x}\ket{L(x)}+U
 \left(\sum_{y,z,v}\beta(y,z,v)
 \ket{y}\ket{z}\ket{v}(\ket{y}\ket{z}-\ket{x}\ket{L(x)})\right)}\quad\\
&=&\!\!\!\! \ket{x}\ket{B}\ket{B}\ket{x}\ket{L(x)}+
 \sum_{y,z}U'\left(\sum_{v}\beta(y,z,v)\ket{y}\ket{z}\ket{v}\right)
 \otimes(\ket{y}\ket{z}-\ket{x}\ket{L(x)}).
\end{eqnarray*}
Since we have $(y,z)\neq (({\rm T}[x],{\rm T}[0]),({\rm T}[x],{\rm T}[1]))$, 
if we write 
$$
U\left(\sum_{y,z,v}\be(y,z,v)\ket{y}\ket{z}\ket{v}
(\ket{y}\ket{z}-\ket{x}\ket{L(x)})\right)
=\ga\ket{x}\ket{B}\ket{B}\ket{x}\ket{L(x)}+\ket{\psi},
$$
then $\l \psi|x,B,B,x,{\rm T}[1-L(x)]\r=0$, 
where $\ket{x,B,B,x,{\rm T}[1-L(x)]}$ denotes 
$\ket{x}\ket{B}\ket{B}\ket{x}$ $\ket{{\rm T}[1-L(x)]}$. 
Moreover, we can see that
$$ 
||\ket{\psi}||^2
\le \left|\left|U\left(\sum_{y,z,v}\be(y,z,v)
\ket{y}\ket{z}\ket{v}(\ket{y}\ket{z}-\ket{x}\ket{L(x)})\right)
\right|\right|^2
\le\sum_{y,z,v}2|\be(y,z,v)|^2=1/2^l. 
$$
Steps 5--7 are implementable by using the symbol strings 
written on the seventh track and the branching lemma \cite{BV97}. 
From the above, $M$ satisfies the statement of this theorem. 
\end{Proof} 

Let $M=(Q,\Si,\de)$ be a QTM and 
$M'=(Q',\Si_1\times\Si_2\times\Si_3,\de')$ be a QTM 
such that $Q\times\Si\subseteq Q'\times(\Si_1\times\Si_2)$. 
We say that $M'$ {\em carries out} $M$ 
{\em with zero error and slowdown $f$}, 
if there exists a function $f:\N^2\rightarrow\N$ 
such that for any input $y=(x,1^l)$ of $M$, where $l\in\N$, 
and any $C\in\cD(M,y)$ there exists some $T'\in\Si^\#_3$ 
(depending on $y$) satisfying the following condition: 
If $M_{\de'}^{f(|x|,l)}\ket{C}\ket{T'}=\ket{\ph}+\ket{\ps_1}$ 
and $E(\hat{T}^2={\rm T}[\star])(\ket{\ph}+\ket{\ps_1})=\ket{\ps_1}$, 
then $(M_\de\ket{C})\otimes\ket{T'}=\ket{\ph}+\ket{\ps_2}$, 
$||\ket{\ps_1}-\ket{\ps_2}||\le 1/2^l$, 
and $E(\hat{T}^2={\rm T}[\star])\ket{\ps_2}=0$. 
Using Theorem \ref{th:44} we can show the following lemma 
by a way similar to the proof of Theorem \ref{th:34}.

\begin{Lemma}\label{th:45} 
If $L$ is in ${\bf ZQP}$, there are a polynomial $p(n,l)$ and 
a polynomial time QTM $M$ such that $M$ carries out 
a polynomial time oracle QTM $M^L$ with zero error and slowdown $p$.
\end{Lemma}

We shall show that ${\bf ZQP}$ is robust by using Lemma \ref{th:45}. 
To this end, we need to construct our algorithm 
so that we cannot erase the symbol $\star$ 
written as a witness of an error in the subsequent steps.

\begin{Theorem}\label{th:46} 
${\bf ZQP}^{\bf ZQP}={\bf ZQP}$.
\end{Theorem}

\begin{Proof} 
Let $L\in {\bf ZQP}^{\bf ZQP}$. 
Then there is a language $L'\in{\bf ZQP}$ such that 
$L\in {\bf ZQP}^{L'}$. We can assume that an oracle QTM $M_1^{L'}$ 
recognizes $L$ with probability $1-1/2^l$. 
Let the computation times of $M_1^{L'}$ be $p_l(n)=p(n,l)$. 
By Lemma \ref{th:45} there are a polynomial $f(n,l)$ and 
a polynomial time QTM $M'$ that 
carries out a polynomial time oracle QTM $M_1^{L'}$ 
with zero error and slowdown $f$. 
Now we consider a {\bf ZQP}-machine $M_l$ 
which implements the following algorithm. 
We assume that the length of the input $x$ of $M_l$ is $n$.

Step 1. $M_l$ writes $1^l$ and $1^{p_l(n)}$ 
on the second and third tapes.

Step 2. $M_l$ repeats the following operation $p_l(n)$ times: 
If the special symbol $\star$ is written 
in the cell 0 of the first tape, 
then $M_l$ changes the string $0^m1^{p_l(n)-m}$ on the third tape 
to $0^{m+1}1^{p_l(n)-m-1}$ in $g(n,l)$ steps. 
Otherwise, $M_l$ carries out $f(n,l)$ steps of $M'$ on $(x,1^l)$ 
in $g(n,l)$ steps. Here, $g$ is an ST-constructible function.

Since the probability that our algorithm incorrectly 
carries out a single step of $M_1^{L'}$ 
and the error probability of $M_1^{L'}$ are both at most $1/2^l$, 
the probability that $M_l$ produces a correct answer 
is at least $(1-1/2^l)^{p_l(n)+1}$. 
Thus, if $1/2^l\le 1/c'(p_l(n)+1)$ with some constant $c'$, 
then $M_l$ recognizes $L$ uniformly larger than $1/2$. 
By Lemma \ref{th:36}, the branching lemma and 
the looping lemma \cite{BV97}, 
step 2 is implementable by a stationary QTM. 
If our algorithm incorrectly carries out a single step of $M_1^{L'}$ 
then the special symbol $\star$, a witness of an error, 
is written on the second track of the first tape by Theorem \ref{th:44}. 
We can see that the construction of our algorithm ensures 
that the symbol $\star$ is not erased in the subsequent steps. 
Therefore, our algorithm is Las Vegas type. 
Now we can choose $l$ such that $1/2^l\le 1/c'(p_l(n)+1)$ 
and that $l$ is a polynomial in $n$, 
and then the computation time of $M_l$ is a polynomial in $n$. 
\end{Proof}


\begin{thebibliography}{999}
\bibitem{BB92}
A. Berthiaume and G. Brassard,\ {The quantum challenge to
 structural complexity theory}, in:\ {\em Proceeding of the 7th Annual
 Structure in Complexity Theory Conference},\ IEEE Computer Society Press,
 Los Alamitos, CA, 1992, pp.132-137. 
\bibitem{DJ92}
D. Deutsch and R. Jozsa,\ {Rapid solution of problems
 by quantum computation},\ {\em Proc.\ Roy.\ Soc.
\ London Ser.\ A}, {\bf 439} (1992), 553-558.
\bibitem{BV97}
E. Bernstein and U. Vazirani,\ {Quantum complexity theory},
 in:\ {\em Proceedings of the 25th Annual ACM Symposium
 on Theory of Computing},\ ACM Press, New York, 1993, pp.\ 11-20.
 Journal version appeared in {\em SIAM J.\ Comput.},\ {\bf 26}
 (1997), 1411-1473.
\bibitem{Sim97}
D. Simon,\ {On the power of quantum computation}, in:\ {\em Proceeding of
 the 35th Annual IEEE Symposium on Foundations of
 Computer Science},\ IEEE Computer Society Press,
 Los Alamitos, CA, 1994, pp.116-123. Journal version appeared in
 {\em SIAM J.\ Comput.},\ {\bf 26} (1997), 1474-1483.
\bibitem{Sho97}
P. W. Shor,\ {Algorithms for quantum computations:
 Discrete log and factoring}, in:\
 {\em Proceedings of the 35th Annual IEEE Symposium on Foundations of
 Computer Science} (S. Goldwasser, ed.),\ IEEE Computer Society Press,
 Los Alamitos, CA, 1994, pp.124-134; {Polynomial-time algorithms 
for prime factorization and discrete logarithms on a quantum computer}, 
{\em SIAM J.\ Comput.},\ {\bf 26} (1997), 1484-1509.
\bibitem{BBBV97}
C. H. Bennett, E. Bernstein, G. Brassard, and U. Vazirani, 
 {Strengths and weaknesses of quantum computing},\
 {\em SIAM J. Comput.}, {\bf 26} (1997), 1510-1523.
\bibitem{AKN98}
D. Aharonov, A. Kitaev, and N. Nisan,\ {Quantum circuits with mixed states},
 in: {\em Proceedings of the 31th Annual ACM Symposium
 on Theory of Computing},\ ACM Press, New York, 1998, pp.20-30.
\bibitem{Yao93}
A. Yao,\ {Quantum circuit complexity}, in: {\em Proceedings of
 the 34th Annual IEEE Symposium on Foundations of
 Computer Science},\ IEEE Computer Society Press,
 Los Alamitos, CA, 1993, pp.352-361.
\bibitem{NO99}
H. Nishimura and M. Ozawa,\ {Computational complexity of
 uniform quantum circuit families and quantum Turing machines},
 {\em Theoret.\ Comput.\ Sci.} (to appear).
\ Available at the LANL quantum physics e-print archive 
at http://xxx.lanl.gov/archive/quant-ph/9906095.
\bibitem{Kit97}
A. Kitaev,\ {Quantum computations: algorithms and error correction},\ 
 {\em Russian Math. Surveys}, {\bf 52} (1997), 1191-1249.
\bibitem{ON98}
M. Ozawa and H. Nishimura,\ {Local transition functions of quantum
 Turing machines}, {\em RAIRO Theor.\ Inform.\ Appl.} (to appear).
\ Available at the LANL quantum physics e-print archive 
at http://xxx.lanl.gov/archive/quant-ph/9811069.
\bibitem{Ben73}
C. H. Bennett, {Logical reversibility of computation},\
 {\em IBM J. Res.\ Develop.},\ {\bf 17} (1973), 525-532.

\end{thebibliography}
\end{document}